\documentclass[aps,prb,twocolumn,amsmath,amssymb,superscriptaddress]{revtex4} 

\usepackage{graphicx}
\usepackage{dcolumn}
\usepackage{bm}

\begin{document}

\title{The Observation of Percolation-Induced 2D Metal-Insulator
  Transition in a Si MOSFET} 

\author{L. A. Tracy}
\affiliation{Sandia National Laboratories, Albuquerque, NM 87185}
\author{E. H. Hwang}
\affiliation{Condensed Matter Theory Center, Department of Physics,
  University of Maryland, College Park, MD 20742-4111} 
\author{K. Eng}
\affiliation{Sandia National Laboratories, Albuquerque, NM 87185}
\author{G. A. Ten Eyck}
\affiliation{Sandia National Laboratories, Albuquerque, NM 87185}
\author{E. P. Nordberg}
\affiliation{Sandia National Laboratories, Albuquerque, NM 87185}
\author{K. Childs}
\affiliation{Sandia National Laboratories, Albuquerque, NM 87185}
\author{M. S. Carroll}
\affiliation{Sandia National Laboratories, Albuquerque, NM 87185}
\author{M. P. Lilly}
\affiliation{Sandia National Laboratories, Albuquerque, NM 87185}
\author{S. Das Sarma}
\affiliation{Condensed Matter Theory Center, Department of Physics,
  University of Maryland, College Park, MD 20742-4111} 

\date{\today}

\begin{abstract}
By analyzing the temperature ($T$) and density ($n$) dependence of the
measured conductivity ($\sigma$) of 2D electrons in the low density ($\sim10^{11}$
cm$^{-2}$) and temperature (0.02 - 10 K)  
regime of high-mobility (1.0 and 1.5 $\times 10^4$ cm$^2$/Vs) Si MOSFETs, we
establish that the putative 2D metal-insulator transition is a density-inhomogeneity driven percolation
transition where the density-dependent conductivity vanishes as
$\sigma (n) \propto (n - n_p)^p$, with the exponent $p \sim 1.2$  
being consistent with a percolation transition.  The `metallic'
behavior of $\sigma (T)$ for $n > n_p$ is shown to be well-described
by a semi-classical Boltzmann theory,  
and we observe the standard weak localization-induced negative
magnetoresistance behavior, as expected in a normal Fermi liquid, in
the metallic phase.
\end{abstract}


\maketitle

The so-called two-dimensional (2D) metal-insulator transition (MIT)
has been a subject \cite{KravchenkoRMP2001,DasSarmaSSC2005} of intense
activity and considerable controversy ever  
since the pioneering experimental discovery \cite{KravchenkoPRB1994}
of the 2D MIT phenomenon in Si MOSFETs by Kravchenko and Pudalov some
fifteen years ago.  The apparent MIT  
has now been observed in almost all existing 2D semiconductor
structures, including Si MOSFETs
\cite{KravchenkoPRB1994,KravchenkoPRB1995}, electrons
\cite{HaneinPRB1998,LillyPRL2003,DasSarmaPRL2005}  
and holes \cite{HaneinPRL1998,SimmonsPRL1998,MillsPRL1999,ManfraPRL2007}
in GaAs/AlGaAs, and electrons in Si/SiGe \cite{LaiPRB2005,LaiPRB2007} 
systems.  The basic phenomenon refers to the observation of a carrier density-induced 
qualitative change in the temperature dependence of the
resistivity $\rho (n, T)$,  
where $n_c$ is a critical density separating an effective `metallic'
phase $(n > n_c)$ from an `insulating' phase $(n < n_c)$, exhibiting
$d\rho/dT > 0$ ($<0$) behavior typical of a metal (insulator). 

The high-density
metallic behavior $(n > n_c)$ often manifests in a large (by $25\%$ for electrons in GaAs/AlGaAs heterostructures
to factors of 2-3 in Si MOSFETs) increase in resistivity 
with increasing temperature in the low temperature (0.05 - 5 K)
regime where phonons should not play much of a role in resistive scattering.  
The insulating regime, at least for very low $(n \ll n_c)$ densities and temperatures, seems to be the conventional 
activated transport regime of a strongly localized system.  The 2D MIT phenomenon occurs in relatively high-mobility systems 
although the mobility values range from $10^4$ cm$^2$/Vs 
(Si MOSFET) to $10^7$ cm$^2$/Vs (GaAs/AlGaAs) depending on the 2D system under consideration.  
The 2D MIT phenomenon is also considered to be a low-density phenomenon although,
depending on the 2D system under consideration, the critical density $n_c$ differs by two orders of magnitude 
($n_c \sim 10^{11}$ cm$^{-2}$ in 2D Si and $\sim 10^9$ cm$^{-2}$ in the high-mobility 
GaAs/AlGaAs heterostructures.  The universal features of the 2D MIT phenomenon are:  (1) the existence of a critical density $n_c$ 
distinguishing an effective high-density metallic ($d\rho/dT > 0$ for $n > n_c$) phase
from an effective low-density insulating ($d\rho/dT < 0$ for $n < n_c$) phase; and (2) while the insulating phase 
for $n < n_c$ seems mostly to manifest the conventional activated transport behavior,
the metallic temperature dependence at low ($n \gtrsim n_c$) densities is universal in the sense that it 
manifests a very strong temperature dependence, not seen in standard 3D metals where
$\rho (T)$ is temperature independent in the $T < 10$ K Bloch-Gr\"{u}neisen transport regime.

The excitement and controversy in the subject arise from the deep
conceptual questions associated with the nature of the MIT and the
metallic phase.  In particular, it is theoretically well-established \cite{KravchenkoRMP2001}
that a non-interacting (or weakly interacting) 
disordered 2D electron system is an insulator at $T = 0$, and
therefore it follows \cite{KravchenkoRMP2001} that if the 2D MIT
is a true $T = 0$ quantum phase transition (QPT), 
then the 2D metallic phase, if it survives all the way to $T = 0$,
must necessarily be a novel and exotic non-Fermi liquid phase since it
cannot be connected adiabatically to the 
non-interacting 2D electron system, which is always localized in the
presence of any disorder.  The alternative possibility
\cite{DasSarmaSSC2005} is that the metallic phase is allowed only at
finite temperatures,  
and the 2D MIT is not a QPT, but a density-induced crossover from a
weakly localized `effective' metallic phase to a strongly localized
insulator.  These two qualitatively distinct viewpoints have 
both received support in the literature, and experimental results have
been presented claiming support for the QPT \cite{AnissimovaNP2007}
and the finite-temperature crossover
\cite{DasSarmaPRL2005,ManfraPRL2007} viewpoints, 
respectively. 
Whether the observed 2D MIT is a novel ($T = 0$) QPT
leading to an exotic 2D non-Fermi liquid metallic phase established by
interaction effects or is a finite-temperature 
crossover between an effective metallic phase and the insulating phase
is obviously a question of great fundamental importance. 

Previous measurements of 2D MIT phenomena in Si MOSFETs have been discussed
in terms of quantum critical phenomena \cite{KravchenkoRMP2001,AnissimovaNP2007}.
However, for 2D electrons and holes in high mobility GaAs/AlGaAs heterostructures, it has been shown
that the 2D MIT can be described as a percolation transition \cite{DasSarmaPRL2005,ManfraPRL2007,LeturcqPRL2003}.
Thus, a natural question to ask is whether the 2D MIT in Si MOSFETs is really different
than for GaAs systems, or whether both Si and GaAs 2D systems undergo the same universal percolation-driven transition.
The Si MOSFET is different than GaAs systems in that it has much lower mobility 
and the disorder in Si MOSFETs is thought to be dominated by `short-range' scattering, which is 
potentially different than for other 2D systems, such as in GaAs where `long-range' scattering may play a larger role. 
This point has been recently emphasized in the literature and it has been proposed that the physics behind the 2D MIT 
for Si MOSFETs versus GaAs systems may differ due to differences in mobility and 
range of the disorder potential (short vs. long-range scattering) \cite{AnissimovaNP2007}.

In this article, we provide the first experimental evidence that the 
MIT for Si MOSFETs is a finite-temperature, density
inhomogeneity driven semi-classical percolation transition, 
and is therefore not a QPT.  Our work shows that the 2D MIT is a
universal percolation transition for both GaAs and Si,
in contrast to claims of a QPT in Si \cite{AnissimovaNP2007}.
As mentioned earlier, the 2D MIT in high-mobility GaAs electron and hole systems
has already been experimentally demonstrated 
to be a percolation transition \cite{DasSarmaPRL2005,ManfraPRL2007,LeturcqPRL2003}, 
but our work provides the first
compelling evidence supporting the percolation scenario in Si MOSFETs.
We also establish that the metallic 
phase is a conventional Fermi liquid by showing that (1) the measured
transport properties in the effective metallic phase qualitatively
agree very well with the conclusions of a semi-classical 
Boltzmann transport theory \cite{DasSarmaPRL1999} taking into account
the detailed density and temperature dependence of carrier screening
properties, and (2) the expected negative 
magnetoresistance signature of the conventional weak localization
behavior \cite{LeeRMP1985} is clearly manifest experimentally in the effective metallic
phase of our samples.  

The two samples used in this study are Si MOSFET structures with a
peak mobility of $\sim 1.5 \times 10^4$ cm$^2$/Vs (sample A) and $\sim
1 \times 10^4$ cm$^2$/Vs (sample B).  For both samples, the 2D
electron gas (2DEG) resides at the Si-SiO$_{2}$ gate oxide interface,  
where the SiO$_{2}$ thickness is nominally 35 nm for sample A and 10 nm for 
sample B.  
The gate oxide is thermally grown in dry O${_2}$ at $900^\circ$ C on FZ $(100)$-oriented high-resistivity 
($> 5$ k$\Omega$-cm for sample A and $> 100$ $\Omega$-cm for sample B, at room temperature) p-type Si wafers.  
Ohmic contacts to the 2DEG consist of n$^+$ Si regions
formed by implantation of As; these contacts are not self-aligned, but are instead patterned and implanted 
before the gate oxide growth.  
An n$^+$ polySi gate, patterned to form a 15 $\mu$m $\times$ 100 $\mu$m Hall bar, is used to induce carriers.
The 2DEG resistivity is experimentally determined via standard
four-terminal lock-in measurements and the density is calibrated via
measurements of the Hall resistivity. 

\begin{figure}
\includegraphics{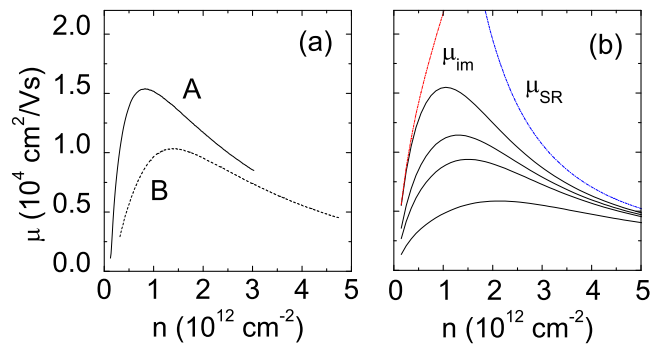}
\caption{\label{fig1} (color online) (a) Experimental mobility $\mu$ as a function of
  density $n$ for two samples, A and B, at a temperature $T = 0.25$
  K. (b) Theoretical mobility as a function of density. High (low)
  density mobility is limited by the surface roughness $\mu_{\rm SR}$
  (charged impurity $\mu_{\rm im}$) scattering. Different curves
  correspond to different charged impurity densities ($n_i = 6, 9, 12$, and $24 \times 10^{10}$ cm$^{-2}$ 
  from top to bottom), but with the same surface roughness ($\Delta = 1.9$ \AA, $\Lambda = 26$ \AA).} 
\end{figure}

In Fig. 1 we show the density-dependent mobility of samples A and B
along with the theoretically calculated mobility curves.  
Our theoretical calculations \cite{comment} include the temperature-dependent finite
wavevector screening of the charged impurity Coulomb scattering and the short-range surface
roughness scattering at the Si-SiO$_2$ interface.
It is well-known that the long-range charged impurity scattering and the short-range surface 
roughness scattering dominate respectively the low and the high carrier density regimes of transport 
in Si MOSFETs \cite{AFS}.
We show Fig. 1a and b together to allow the reader to
judge the level of agreement between experiment and theory, which depends on several
scattering parameters (e.g. the density and spatial distribution of charged impurity 
centers in the sample and the height and lateral correlation functions of the interface roughness function).
The mobility ($\mu$) first rises steeply with density ($n$) reaching a
maximum at a characteristic density $n_m$, beyond which 
$\mu$ falls off slowly with increasing $n$, due to  short-range surface roughness 
scattering at the Si-SiO$_2$ interface \cite{AFS}.
At low carrier densities ($n < 10^{12}$ cm$^{-2}$) however, long-range  
Coulomb scattering by unintentional random charged impurities invariably
present on the 
insulating oxide side of the Si-SiO$_2$ interface dominates the 2D
mobility, and, as has been emphasized \cite{KlapwijkSSC1999} in the
literature, all Si 2D MIT behavior manifests itself at low carrier  
densities ($n \sim 10^{11}$ cm$^{-2}$) where the charged impurity
scattering dominates transport and the short-range interface roughness
scattering is negligible.  
We emphasize the fact that the 2D MIT
phenomenon (i.e. $n_c$ and the density range for the 
`anomalous' metallic phase with a strongly temperature dependent 2D
resistivity) occurs in the density regime $n \sim 10^{11}$ cm$^{-2}$
which, being substantially below $n_m$, is completely dominated by the long-range screened
Coulomb scattering.  
This dominance of Coulomb scattering in the 2D
MIT physics is crucial for 
the percolation physics which arises from the nonlinear failure of
screening of the long-range Coulombic charged impurity potential, to
be discussed below.
The short-range surface-roughness scattering, which dominates MOSFET transport for 
$n > 10^{12}$ cm$^{-2}$, plays no role in the 2D MIT phenomena since both the critical density 
for the transition and the density regime where the strong metallic behavior 
(i.e. the strong temperature dependence of the 2D conductivity) is observed are well below the 
operational regime of the short-range surface roughness sacattering.

\begin{figure}
\includegraphics{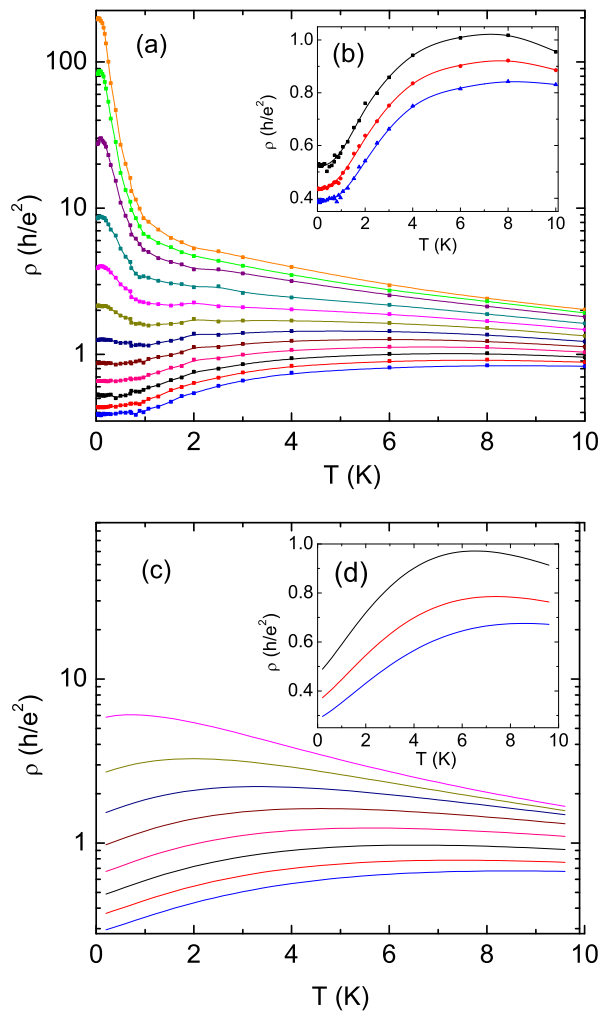}
\caption{\label{fig2}  (color online) (a) Experimental resistivity $\rho$ in units of
  $h/e^2$ as a function of temperature for sample A at 2D electron
  densities (from top to bottom) $n = 1.07, 1.10, 1.13, 1.20, 1.26,  
1.32, 1.38, 1.44, 1.50, 1.56, 1.62$, and $1.68 \times 10^{11}$ cm$^{-2}$.
Inset (b)  Enlarged view of data in the effective metallic regime at $n = 1.56, 1.62$, and $1.68 \times 10^{11}$ cm$^{-2}$.
Solid lines are guides to the eye. (Data below 0.1K may not be
reliable due to electron heating.) (c) Theoretically
calculated temperature and density dependent resistivity for sample A
for densities $n=1.26, 1.32, 1.38, 1.44, 1.50, 1.56, 1.62$, and
$1.68 \times 10^{11}$ cm$^{-2}$ (from top to bottom). Inset shows the
resistivities for $n=1.56, 1.62$, and $1.68 \times 10^{11}$ cm$^{-2}$.
}
\end{figure}

In Fig. 2 we show our measured (Fig. 2(a) - (b)) and calculated
(Fig. 2(c) - (d)) temperature and density dependent resistivity ($\rho (T,
n)$) for sample A with the maximum mobility of  
$1.5 \times 10^4$ cm$^2$/Vs (see Fig. 1(a)).  The results for sample B
are similar, but with a larger value of the critical density, $n_c \sim 2.5 \times 10^{11}$ cm$^{-2}$
(and somewhat weaker temperature dependence), consistent with its lower 
(and consequently, higher disorder) peak mobility $\sim 1 \times 10^4$ cm$^2$/Vs.   
The classic 2D MIT behavior is apparent in Fig. 2(a) where $\rho (T)$
for various densities manifests the clear distinction between metallic 
($d\rho/dT > 0$) and insulating ($d\rho/dT < 0$) behavior separated
(visually) by a critical density $n_c \sim 1.4 \times 10^{11}$
cm$^{-2}$.  
We note that the temperatures quoted in Fig. 2a and b are
those of the cryostat cold finger; although the resistivity continues to evolve
below $T \sim 100$ mK (albeit slowly), we cannot claim in this regime
($T < 100$ mK) that the 2D electrons and the cold finger are in
thermal equilibrium down to the base temperature of our cryostat (20 mK).
The inset in Fig. 2(b) shows in detail the metallic behavior for
$n > n_c$, which manifests more than a factor of 2 increase in $\rho (T)$
for $T \sim 0.1 - 7$ K  
before decreasing slightly at higher temperatures.
The strong initial increase in $\rho (T)$ with $T$ arises
\cite{DasSarmaPRL1999} from the temperature-induced weakening of
screening at these 
low densities as $T/T_F$ increases from 0.01 at low temperatures
essentially to 1 at $T \sim 8$ K since $T_F \approx 7.3$ K for $n =
10^{11}$ cm$^{-2}$. (Phonon scattering is unimportant in Si MOSFET for
$T<10$ K.) 

Finally, in Fig. 2(c) we show our theoretically calculated $\rho (T,
n)$ where the Boltzmann transport is calculated using the screened 
charged impurity scattering as the only resistive scattering mechanism
\cite{DasSarmaPRL1999}.
Although the ratio of the Coulomb to Fermi energy, $r_s$, can be as large 
as $\sim 15$ near the MIT for this sample, suggesting that Coulomb interactions
could play an important role,
we note that the basic features of the
experimental metallic phase are well-captured by the
screening theory 
with $\rho (T)$ rising approximately by a factor of 2 initially and
then decreasing slightly at higher (lower) temperatures (densities)
around $T/T_F \gtrsim 1$ due to quantum-classical crossover 
\cite{DasSarmaPRL1999}.  
The theoretical results, following the work
of Ref. \cite{DasSarmaPRL1999}, are being shown here in order to
demonstrate that a 
realistic semiclassical transport theory including the screened 
long-range scattering by charged impurities completely captures all
the main features of the experimental results with the basic
physics being a strong temperature and density dependent 
modification in the effective screened disorder due to the large
change in the effective $T/T_F$ in the experimental $T \sim 0.1 - 10$
K temperature range.  
We point out that, by definition, the 
theoretical transport results in Figs. 2(c) and (d) apply only in the metallic
regime since the metal-insulator transition cannot be captured in a
Boltzmann theory.  

\begin{figure}
\includegraphics{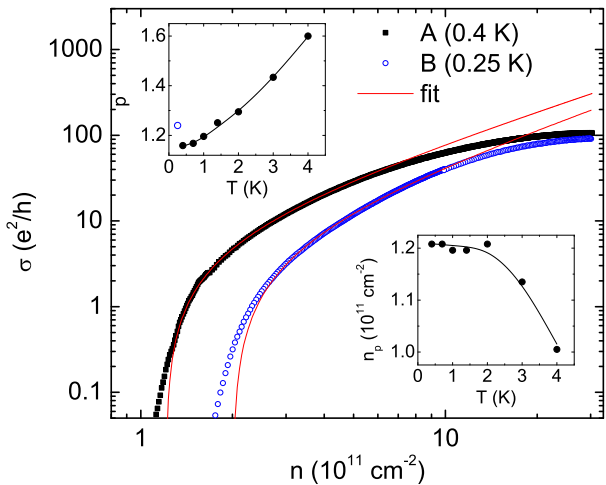}
\caption{\label{fig3} (color online) (a) Main plot:  Points show experimental
 conductivity $\sigma$ in units of $e^2/h$ versus electron density
 $n$ for sample A at $T$ = 0.4 K (closed symbols) and B at $T$ = 0.25 K (open symbols).   
The solid lines are fits to the data of the form
$\sigma = A(n - n_p)^p$.  The upper and lower insets show the exponent
$p$ and critical density $n_p$, respectively, 
 as a function of temperature for sample A.  Solid lines are a guide to the eye.
The open symbol in the upper inset shows $p$ for sample B at $T = 0.25$ K.}
\end{figure}

Having established the phenomenology of the 2D MIT behavior in Fig. 2,
we now come to the nature of the density driven transition itself.  We
plot in Fig. 3 our measured conductivity $\sigma (n)$ as a  
function of $n$ to see if a percolation behavior, $\sigma (n) \sim (n
- n_p)^p$, where $n_p$, $p$ are the percolation 
transition density and exponent respectively, is manifested.  
Our fit yields $p \approx 1.20$, $n_p \approx 1.2 \times
10^{11}$ cm$^{-2}$ for sample A and $p \approx 1.24$, $n_p \approx 2.0 \times
10^{11}$ cm$^{-2}$ for sample B.   The fits describe  
$\sigma (n)$ well over nearly two orders of magnitude change in conductivity.
The deviations of the fit from the data seen at the lowest densities ($n \lesssim n_c$) are 
expected since our percolation analysis applies only in the metallic regime. 
At high densities the fit is also expected to deviate since
the percolation fit is a critical scaling behavior which is expected to be best for densities close to $n_p$,
and also the role of surface roughness scattering becomes increasingly important at high densities.
The fact that our best fit values of $n_p$ and $p$ for sample A are almost
independent of temperature at low temperatures 
is a good indicator for the percolation transition being the correct
description for the 2D MIT phenomenon shown in Fig. 2.  
Our numerically extracted best-fit exponent $p \approx 1.2$ is sufficiently
close to the percolation conductivity exponent of 1.31 for us to feel confident
that the observed 2D MIT in Si MOSFETs is a density inhomogeneity-driven
percolation transition.

\begin{figure}
\includegraphics{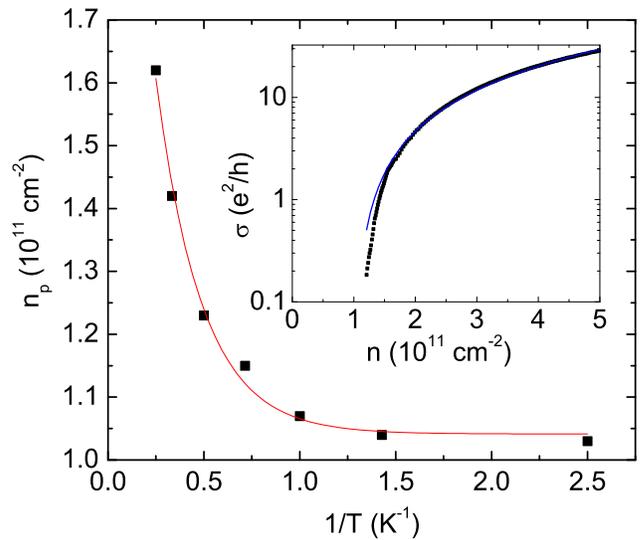}
\caption{\label{fig4} (color online) Main plot:  Points show $n_p$ vs. $1/T$
obtained from a percolation fit to the $\sigma$ vs. $n$ data performed by holding the
exponent $p = 1.31$ fixed (see text).  The solid trace is a fit of the form $n_p = n_0 + Ae^{-b/T}$.
Inset:  Points show experimental data $\sigma$ vs. $n$ for sample A at $T = 0.4$ K, while
the solid line is a percolation fit with fixed $p = 1.31$.}
\end{figure}

In Fig. 4, we show an alternate (and perhaps more appropriate, if the transition is indeed a percolation transition)
method of extracting the critical density from a percolation fit,
where we hold both the exponent $p$ 
and the prefactor $A$ fixed, and allow only $n_p$ to vary.  We use
$p = 1.31$, the value expected for percolation in 2D,
and determine $A$ from the percolation fit to the $T = 0.4$ K data, as shown
in the inset to Fig. 4, also holding $p = 1.31$ fixed. 
If the 2D MIT is indeed a percolation transition, then this method of obtaining $n_p$ is more
appropriate than allowing $p$ to vary since the exponent is universal and the 
critical density is not.  As seen in the inset to Fig. 4, the fit to our $\sigma$ vs. $n$ data
with $p = 1.31$ fixed is very good.
As can be seen from the main figure, the extracted $n_p$ obtained via this method decreases
with decreasing temperature, eventually saturating at the lowest temperatures.  A fit of the form
$n_p = n_0 +Ae^{-b/T}$ is shown as the solid line in the main plot of Fig. 4.  The fit yields
$n_p = 1.04 + 1.62e^{-4.20/T}$.
The parameter $b = 4.2$ K is an effective "energy gap" that can be calculated if the impurity
distribution is known.
We note that this implies a $T=0$ MIT transition density of $1.04 \times 10^{11}$ cm$^{-2}$, which is considerably 
less than the nominal critical density of $1.4 \times 10^{11}$ cm$^{-2}$ one would have inferred from the 
temperature-dependent resistivity behavior in Fig. 2(a).  The fact that the temperature-dependent resistivity 
by itself gives a strong over-estimate of the 2D MIT critical density is already known in the literature.

In Fig. 5 we show our observed weak localization behavior by plotting
the expected \cite{LeeRMP1985} negative magnetoresistance in the
metallic phase as a function of an applied weak magnetic 
field.  The weak localization data in Fig. 5 is fitted to the standard
di-gamma function behavior expected of a disordered 2D
system\cite{LeeRMP1985} 
\begin{equation}
\Delta\sigma = -\alpha\frac{e^2}{h}\frac{g_v}{\pi}\left[ \Psi \left(
    \frac{1}{2} + \frac{\tau_B}{\tau} \right) - \Psi \left(
    \frac{1}{2} + \frac{\tau_B}{\tau_\phi} \right) \right], 
\end{equation}
where $\tau_B \equiv \hbar/4eB_{\bot}D$ and $D$ is the diffusion
coefficient.  The behavior is quite normal, explicitly demonstrating
that our observed metallic phase ($n > n_c, n_p$) is indeed the  
usual weakly localized Fermi liquid metallic phase, and not some
exotic non-Fermi liquid $T = 0$ 
metal.  The observation of the expected weak localiation behavior in
the 2D metallic phase along with the percolative nature of the
transition is strong evidence that the experimental 2D MIT 
is not a QPT, but a crossover.

\begin{figure}
\includegraphics{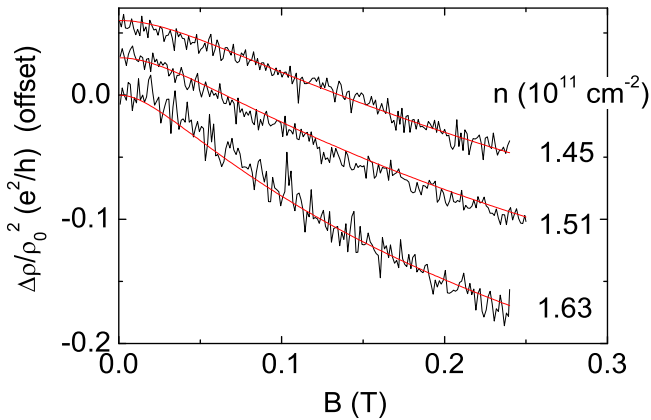}
\caption{\label{fig5} (color online) Measured weak localization correction to the
  resistivity, $\Delta \rho_{xx}/\rho_0^2$ in units of $e^2/h$ (with offset for clarity) versus
  perpendicular magnetic field $B$ at densities $n = 1.51$, 1.45, and $1.63 \times
  10^{11}$ cm$^{-2}$ 
at $T = 100$ mK.  The red lines are fits to the standard weak
localization correction (see text).  The momentum relaxation times
$\tau$ are determined from the mobility, while the weak localization
fits yield a correction  
amplitude $\alpha = 0.25$ and phase relaxation time $\tau_{\phi}$ = 33, 32, and 28 ps (top to bottom).}
\end{figure}

Before concluding, we briefly discuss the physics underlying the
percolation transition.  As emphasized above, the 2D
MIT phenomenon occurs in the transport regime dominated by long-range 
Coulombic charged impurity scattering.
It was pointed out some 
years ago \cite{EfrosSSC1989} that such a transport regime is
susceptible at low carrier densities to a semiclassical percolation
MIT since large scale density inhomogeneities 
(i.e. puddles) could appear in the system due to the nonlinear failure
of screening of the charged impurities at low carrier densities.  One
can approximately estimate \cite{EfrosSSC1989} the percolation
transition density, $n_p$, by considering the inhomogeneous
density fluctuations in the system, leading to $n_p \sim 0.2
\sqrt{n_i}/d$, where $n_i$ is the effective 2D charged scatterer
density and $d$ is the effective separation between the 2D carriers
and 
the scatterers.  Using $n_i = 6$ and $11 \times 10^{10}$ cm$^{-2}$, the value needed to get the correct mobility at
higher densities for sample A and B, respectively (these values of $n_i$ are roughly consistent with C-V 
measurements which estimate the density of fixed charge near the Si-SiO$_2$  
interface to be $\sim 8 \times 10^{10}$ cm$^{-2}$), and taking $d \sim 5 - 6$ nm, which is consistent
with the 2D electron layer in the Si MOSFET being about 5 nm 
away from the interface due to the self-consistent potential produced
by the electrons themselves, we get $n_p \sim 0.9$ and $1.2 \times 10^{11}$
cm$^{-2}$ for sample A and B, respectively, which is close to our 
experimentally extracted percolation density in Fig. 3(a).  (We note
that $d$ corresponds to the effective separation between the 2D
electrons and the charged impurities, which must incorporate the fact
that the 2D electrons are on the average about 5 nm inside Si at $n
\sim 10^{11}$ cm$^{-2}$.)  
The 2D conductivity percolation exponent is numerically known to be
around 1.31, which is close to the exponent value (1.20 for sample A, 1.24 for sample B) 
we get in our analysis.  
Finally, we note that the puddles produced by
the nonlinear failure of screening of charged 
impurities in low-density 2D systems have been directly observed in 2D
GaAs \cite{IlaniPRL2000} systems. 

In conclusion, we have experimentally established that the 2D MIT in
Si MOSFETs is a percolation-induced transition driven by the dominance
of the long-range charged impurity disorder.  
At low 
carrier densities, the failure of linear screening leads to the
formation of inhomogeneous puddles in the 2D density landscape, which
then produces a semi-classical percolation transition.  The nature 
of the percolation transition and the effective metallic phase is the
same in Si MOSFETs and 2D GaAs electron \cite{DasSarmaPRL2005} and
hole \cite{ManfraPRL2007} systems.  
The metallic 
temperature dependence arises from the strong temperature dependence
of screening and the percolation transition arises from the
low-density failure of screening.  The 2D MIT is therefore a 
quantum crossover phenomenon, not a QPT, from an effective (weakly
localized) metallic phase to a strongly localized insulating phase.  
The agreement between the percolative
conductivity and our experimental work as well as our finding of the usual weak
localization effect in our data conclusively rules out quantum criticality in the
2D MIT phenomena.  Our work establishes that the 2D MIT is a universal
percolation-driven transition, caused by the inhomogeneous density landscape
induced by charged impurities at low carrier densities, in both GaAs-based systems
\cite{DasSarmaPRL2005,ManfraPRL2007,LeturcqPRL2003}
as well as in Si MOSFETs (our work).
Screening of the charged impurity scattering produces the strong temperature dependence 
of the apparent metallic phase since low carrier densities automatically imply large values of 
$T/T_F$, and the eventual failure of screening at still lower densities, where there are just too 
few carriers to effectively screen the impurity charges, leads to the percolation MIT since the 
2D system becomes inhomogeneous due to the formation of puddles.

\begin{acknowledgments}
This work was performed, in part, at the Center for Integrated Nanotechnologies, 
a U.S. DOE, Office of Basic Energy Sciences user facility. 
Sandia National Laboratories is a multi-program laboratory operated by Sandia Corporation, 
a Lockheed-Martin Company, for the U. S. Department of Energy under 
Contract No. DE-AC04-94AL85000.
\end{acknowledgments}


\begin{thebibliography} {99}
\bibitem{comment} All calculations are carried out following the
  prescription described in refs. 16 above, where all the theoretical
  details are available.
\end{thebibliography}
\end{document}